\documentclass{sf2a-conf2012}
\usepackage{graphicx}
\usepackage{hyperref}
\usepackage[]{natbib}  
\usepackage[cyr]{aeguill}
\usepackage{epstopdf}

\def\BibTeX{{\rm B\kern-.05em{\sc i\kern-.025em b}\kern-.08em
    T\kern-.1667em\lower.7ex\hbox{E}\kern-.125emX}}
\bibpunct{(}{)}{;}{a}{}{,}  
\newcommand{\hi}{\mbox{H{\sc i}}}
\newcommand{\kms}{km~s$^{-1}$}
\usepackage{amssymb,amsmath}

\begin{document}

\TitreGlobal{SF2A 2012}


\title{Know (better) your neighbour:\\
 New \hi\ structures in Messier 33 unveiled  by a multiple peak analysis of high-resolution 21-cm data}
\runningtitle{\hi\ kinematics, dynamics and structure of Messier 33}
\author{Laurent Chemin}\address{LAB, CNRS UMR 5804,  Universit\'e de Bordeaux, F-33270, Floirac, France, \textbf{contact:  chemin@obs.u-bordeaux1.fr}}
\author{Claude Carignan}\address{Dept. of Astronomy, University of Cape Town, Rondebosch 7700, South Africa}
\author{Tyler Foster}\address{Dept. of Physics and Astronomy, Brandon University, Brandon, MB R7A 6A9, Canada}
\author{Zacharie Sie Kam}\address{D\'ept. de physique, Universit\'e de Montr\'eal, Montr\'eal, QC H3C 3J7, Canada}



\maketitle


\begin{abstract}
In our quest to constrain the dynamical and structural properties of Local Group spirals
from high-quality interferometric data, we have performed a neutral hydrogen survey in the direction 
of Messier 33. Here we present a few preliminary results from the survey and show the benefits 
of fitting the \hi\ spectra by multiple peaks on constraining the structure of the Messier 33 disk.
In particular we report on the discovery of new  inner spiral-like and outer annular structures overlaying 
with the well-known main \hi\ disk of Messier 33. Possible origins of the additional outer annular structure are presented.

\end{abstract}



\section{Context}

 The dynamical and structural properties of \hi\ disks of nearby spirals 
 mainly result from the analysis of the 0th and 1st moments of \hi\ spectra obtained 
 from single-dish and interferometric observations. 
 Curiously more thorough analyses of \hi\ spectra making profit from current high spectroscopic precision and sensitive cm-data
 remain rare.

 In 2006 we have started a  \hi\  survey of the most massive spiral disks from the Local Group 
 (except the Milky Way) to revisit their structure, kinematics and dynamics. 
 Aperture synthesis at DRAO combined with short spacing data have been used to perform 21-cm observations of the Andromeda galaxy (Messier 31)
 at spectral resolution $\lesssim 5$ \kms, angular resolution $\sim 300$ pc ($D \sim 800$ kpc) and sensitivity 
 down to $\sim 2 \times 10^{19}$ cm$^{-2}$ \citep{che09}.

 Since many spectra are far from being dominated by one single \hi\ component 
 we have shown  that the moment analysis of datacubes was not appropriate  
 \citep[see Fig. 1 and \S 3.2 of][and left panel of Fig.~\ref{fig:fig1} below]{che09}. 
  This is the reason why we developed a `search and fit' algorithm of multiple (gaussian) components. 
  Applied to new Messier 31 data this algorithm has allowed the detection of sometimes up to five \hi\ significant components per profile, which 
  had never been reported beforehand for nearby \hi\ spirals. 
  So many multiple peaks likely result from the combination of extreme projection effects of the warped Messier 31 disk with internal 
  and external dynamical perturbations (spiral density wave, lagging halo, expanding gas shells,
  accretion of gas from the intergalactic medium or from nearby minor companions, etc). 
  The discovery of outer \hi\ spurs and spiral arm  was also reported, as well as the characterization of the disk warp in terms of twist and tilt 
  angles and the measurement of the most extended rotation curve for Messier 31.  
  
   We note that this kind of hyperspectral decomposition  within multiple gaussian peaks 
  is not new and has been used several times \citep[e.g.][]{sic97,oh08}. It is nonetheless not generalized in \hi\ studies. 
   From a dynamical point-of-view, the multiple peak analysis has led  
  to (marginally) different rotation velocities and inclinations than those derived with another recent and high-quality 
   \hi\ datacube of Messier 31 from the 0th- and 1st-moment analysis \citep{bra09,cor10}. 
   Again, such differences have already been reported  \citep[see e.g. Figs. 11 and 13 of][]{oh08}.   

\section{Yet another new \hi\ survey of Messier 33}
  
   In pursuit of our project we present here very preliminary results for Messier 33, a late-type spiral whose \hi\ disk 
   is known to be  warped \citep{cor97}.  
   The 21-cm interferometric data were still obtained at DRAO \citep[combined with the Arecibo data of][]{put09} 
   but at a  larger spectral resolution (3.3 \kms) than for Messier 31 observations. 
   Of course it is very likely to detect multiple components with highly resolved  spectra. However 
   this  does not guarantee the success of detecting \textit{realistic} ones because noise 
   becomes important at  high resolution.  Furthermore the 
   number of components that can be fitted per spectra depends on the resolution. With more and more  peaks found in 
   an individual spectrum (as for Messier 31), it becomes less and less straightforward to interpret the data and 
   identify for instance the component that is the most representative of the disk 
   circular rotation to those that are caused by all abovementioned perturbing effects. 
   The \hi\ datacube of Messier 33 has thus been filtered to lower resolution  to simplify the hyperspectral 
   decomposition.

\section{Preliminary results: evidence for new \hi\ structures in Messier 33}

Other recent \hi\ surveys of Messier 33 have been performed at VLA and Arecibo \citep{thi02,put09}. 
The  VLA data of \citet{thi02} have allowed to determine  for the first time the inner structure of the \hi\ disk with unprecedented details (resolution of 20 pc).  
The Arecibo data of \citet{put09} were more appropriate to study the nearby environment of Messier 33 at a resolution of about 1 kpc. 
In particular they have shown the \hi\ disk of Messier 33  is surrounded by arc-like structures and clumps. A hint of such  
perturbations had been presented in another (earlier) Arecibo view of Messier 33 \citep{cor89}. 
Our DRAO survey has thus an intermediate angular resolution to them. 

\begin{figure}[ht!]
 \centering
 \includegraphics[height=4.4cm]{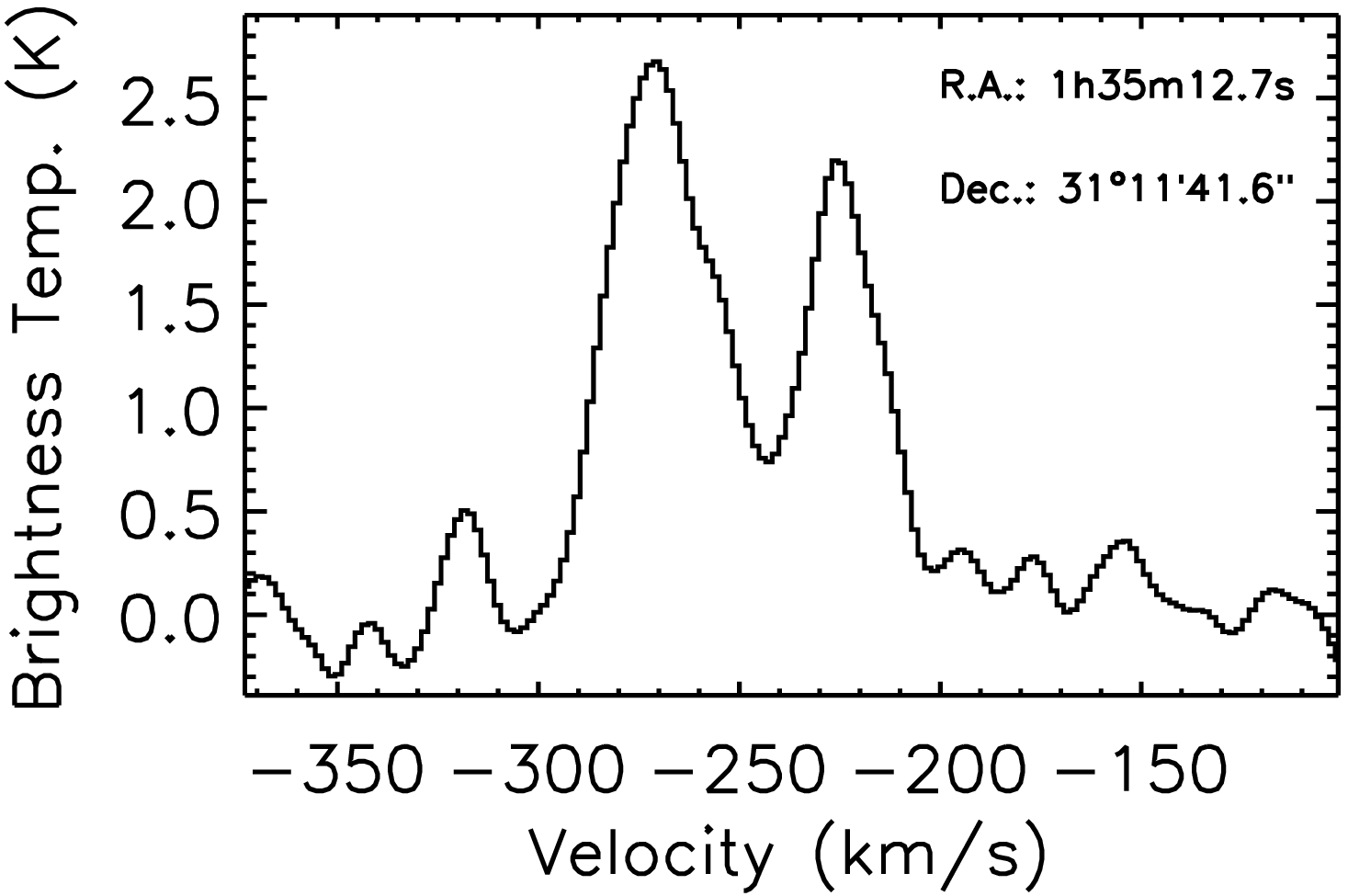}\includegraphics[height=4.4cm]{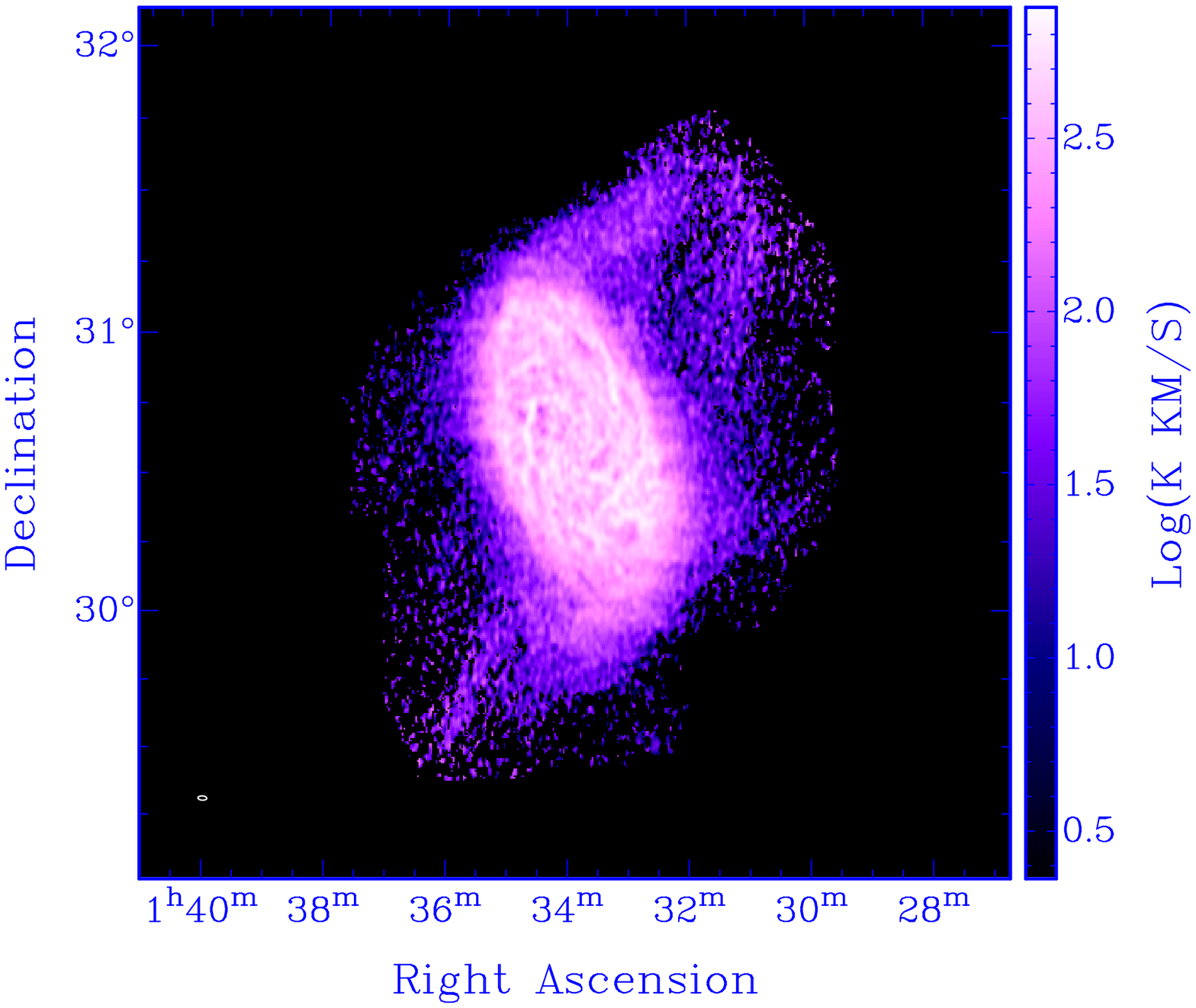}\includegraphics[height=4.4cm]{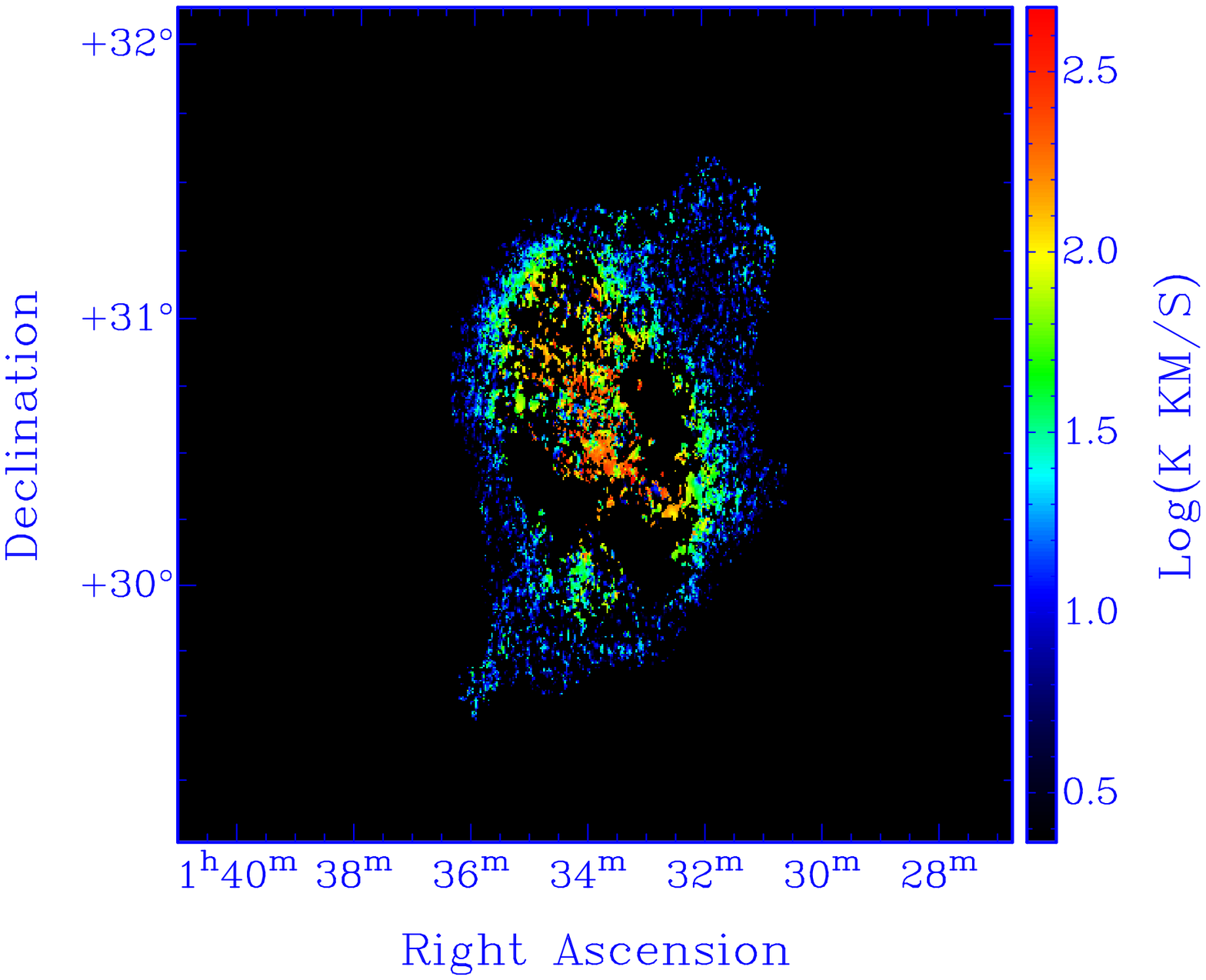}     
  \caption{Illustration of a \hi\ profile with two distinct components (left). Total integrated \hi\ emission map of Messier 33 (middle). 
  \hi\ integrated emission map of the `secondary' fitted \hi\ component in Messier 33 (right). A logarithmic stretch is used for them. The
   ellipse in the bottom-left corner of the midle panel represents the size of the synthesized beam.}
  \label{fig:fig1}
\end{figure}

Working with a datacube of effective spectral resolution of 10 \kms\  our `search and fit' algorithm of multiple peaks 
identifies sometimes up to 3 significant \hi\ components in the datacube.  An example of two distinct components is shown 
in Fig~\ref{fig:fig1} (left-hand panel). Here the   components are separated by $\sim$ 45 \kms.  
The total integrated \hi\ emission of Messier 33 is shown in Fig~\ref{fig:fig1} (central panel). 
The external arc-lile structure and the SW clump are clearly detected, even within our $\sim 300$-pc resolution data, as well   
as the `main' inner disk. 
Multiple components are not observed over the whole field-of-view, as seen in Figure~\ref{fig:fig1} (right panel), but are 
preferentially distributed along a `secondary' spiral-like structure in the inner disk  and 
an annular structure in the outer regions ($r \sim 80'$ or 19 kpc).  It is obvious that none of these new structures would have been 
 identified with a moment analysis of the datacube.
  
A preliminary tilted-ring model has been fitted to the velocity field of the `main' \hi\ component shown in the left-hand panel 
of Fig.~\ref{fig:fig2}. A significant twist of the orientation of the major kinematical axis is evidenced, 
as well as a tilt of the \hi\ disk (Fig.~\ref{fig:fig3}).  
This result thus confirms the warped nature of the \hi\ disk of Messier 33.  
The kinematics of the external arc-like structure does not differ so much from that of the inner disk, implying 
that this perturbation is bound to the disk. We have not yet fitted the warp parameters for it, as shown by constant 
inclination and position angles at those locations ($r > 100'$, Fig.~\ref{fig:fig3}).

   The kinematics of the outer annular structure is shown in the middle panel 
   of Fig.~\ref{fig:fig2} and its residual field when subtracted from the velocity field of the `main' \hi\ component 
    in  the right-hand panel of Fig.~\ref{fig:fig2}. 
    Differences of radial velocities   sometimes reach 40-50 \kms\ in absolute values. 
    At this stage of our analysis it is too early to firmly identify 
    which of the multiple components is the real tracer of the `main' disk kinematics to that of the inner `secondary' spiral-like structure on one hand,  
    and to that of the outer \hi\ annulus on another hand. Indeed the disk kinematics is strongly 
    perturbed in those regions (warp,  connection with the external arc-like structure, etc).        
    It is also too early to constrain the exact origins of the inner `secondary' spiral-like pattern and 
     the outer annular structure.  Origins for this later could be:
 
 \begin{figure}[t!]
 \centering
 \includegraphics[height=4.4cm]{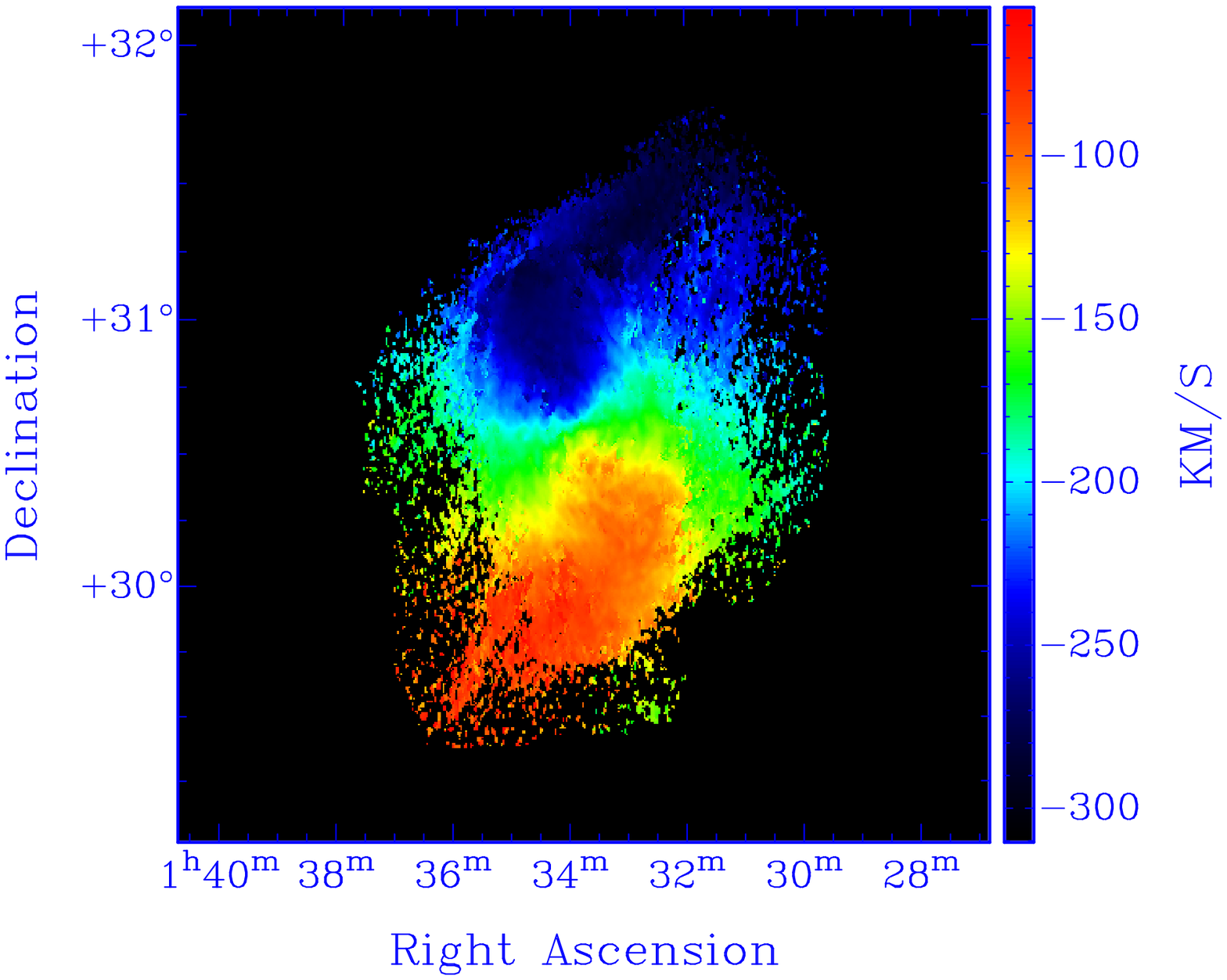}\includegraphics[height=4.4cm]{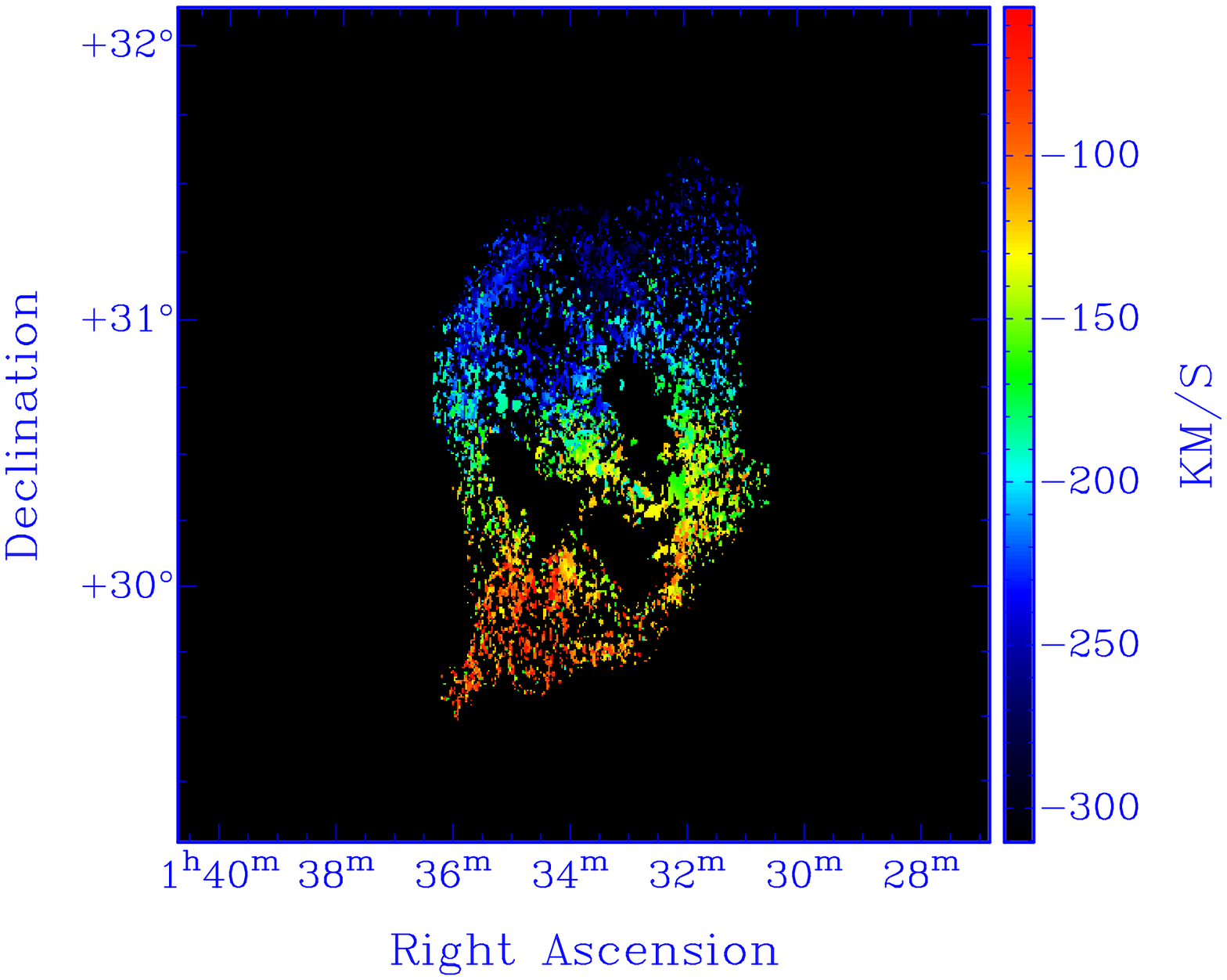}\includegraphics[height=4.4cm]{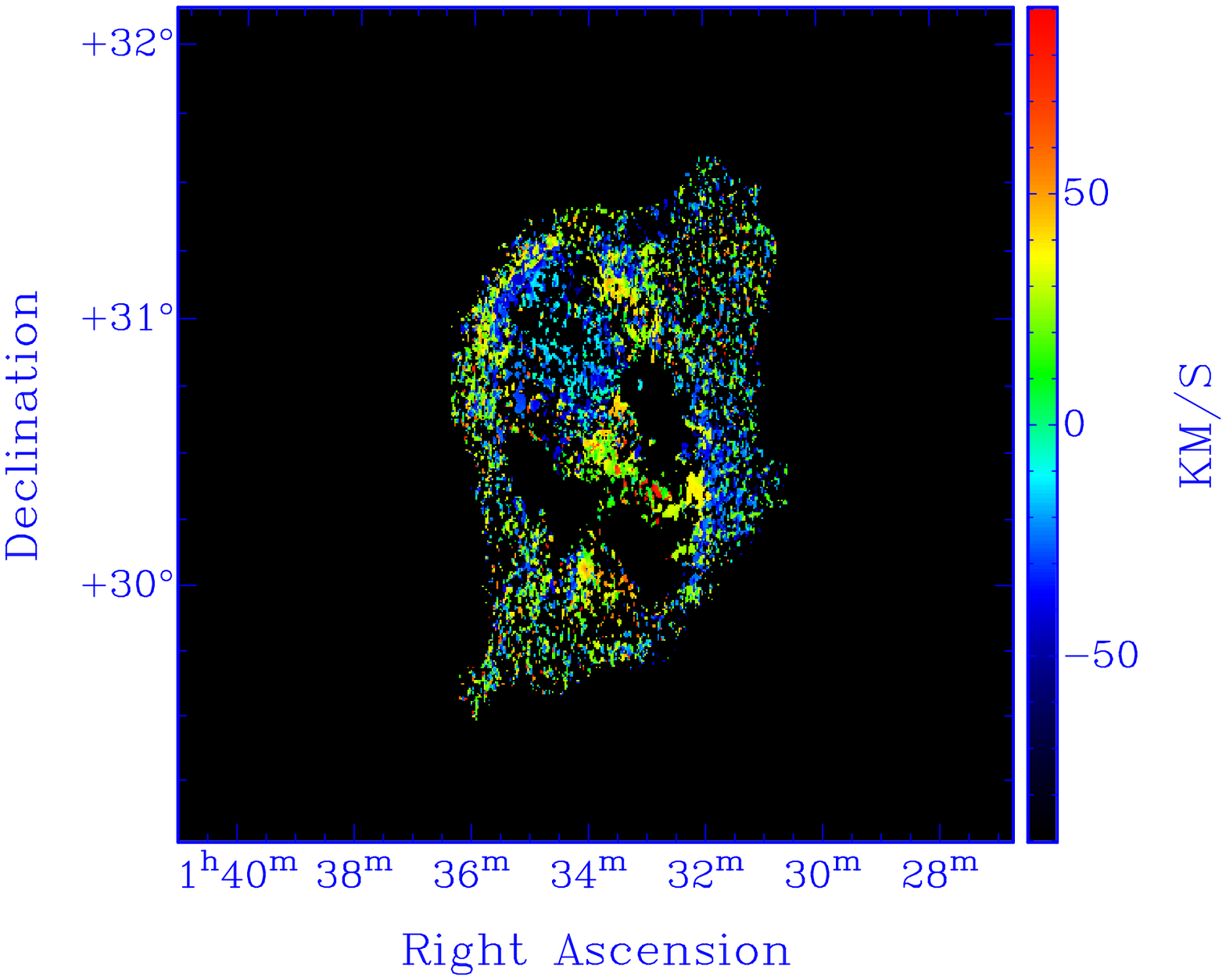}      
  \caption{Velocity fields of the `main' \hi\ component of Messier 33 (left) and of the secondary component (middle). The residual 
  field from their mutual subtraction is shown in the right-hand panel.}
  \label{fig:fig2}
\end{figure}
 
 \begin{itemize}
  \item  The   annular structure has external origins to Messier 33. Gas accretion on the outer disk parts from e.g. the  
 external arc-like strucure or the intergalactic medium could be ongoing. 
 Messier 33 has an obvious  perturbed environment, and past tidal interactions with other galaxies may not be excluded. Numerical
  simulations would be needed to test those assumptions.   
 \item The annular structure is a genuine ring, with internal origins. For instance it could have been developed by gas accumulation at the outer Lindblad resonance. 
  In this case an obvious perturbing density wave could be grand-design spiral structure of Messier 33. This hypothesis could be  tested 
  by measuring the pattern speed of the spiral density wave with a modified version of the Tremaine-Weinberg method,
   and by determining the locations of various Lindblad resonances. 
  \item The annular structure is not a real ring but is only caused by a fortuitous projection effect of a peculiar warping of Messier 33  
  (and maybe also a disk flaring) at the periphery of the \hi\ disk. 
  One would need here gas orbits that have orientation angles significantly different from the constant one displayed in Fig.~\ref{fig:fig3} from $r \sim 85'$ 
  to generate a distinct structure in superimposition to the outer disk.  
 \end{itemize}

Noteworthy is the fact that insights for asymmetric \hi\ profiles  
along a  ring-like structure as caused by the warped gas orbits has been reported in \citet{cor97}. 
  The location of that ring-like structure found by \citet{cor97} corresponds with that of the external arc-like structure, but not to 
  that of the outer \hi\ annulus we evidence here.  Furthermore the \hi\ annulus does not share the same orientation parameters than the external arc-like structure 
  (Fig.~\ref{fig:fig1}).  Two different ring-like structures thus seem to coexist in the outer regions of Messier 33.

\begin{figure}[ht!]
 \centering
 \includegraphics[width=0.5\textwidth]{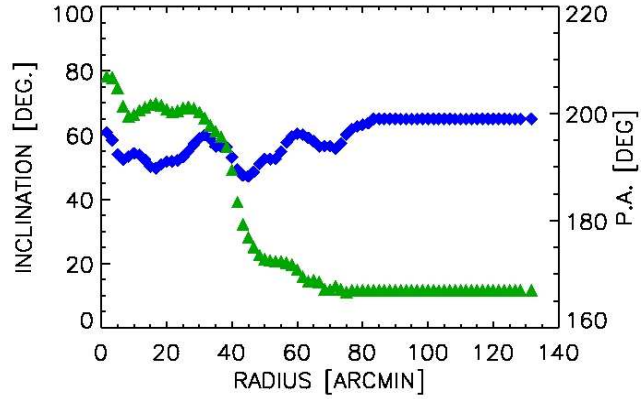}
  \caption{Preliminary results of the tilted-ring model fitted to the `main' \hi\ velocity field of Messier 33 from central panel of Fig.~\ref{fig:fig2}. 
   Blue squares are for the disk inclination and 
      green triangles for the position angle of the major kinematial axis.}
  \label{fig:fig3}
\end{figure}

\section{Conclusions}

Provisional results from a new \hi\ survey in the direction of Messier 33 performed with aperture synthesis observations
 at the Dominion Radio Astrophysical Observatory have been presented. 
 Evidence for new \hi\ structures in Messier 33 have been found from a multiple \hi\ peak analysis of the datacube. Among them is the 
 detection of an annular-like structure in the outer regions of the  \hi\ disk.  That annulus does not correspond to the 
 already known arc-like structure around Messier 33. 
  Complete details of the observing campaign,  the data reduction and  the hyperspectral decomposition    
    will be presented soonly \citep{che13}. Our main objectives are to revisit the structure and dynamics of Messier 33, 
    derive an accurate and  extended rotation curve for it, and model its mass distribution.    
 With the results already obtained for the Andromeda galaxy, this new dataset should help to better constrain the 
 evolution of massive spirals in the Local Group. 
 
\begin{acknowledgements}
We are very grateful to  Mary Putman and Nigel Douglas   for having provided us with their single dish data.
\end{acknowledgements}

\end{document}